\documentstyle[12pt,epsfig,rotating]{article}
\def\bd{
\begin{document}} \def\ed{\end{document}}
\def\bmp{\begin{minipage}} \def\emp{\end{minipage}}
\def\bcc{\begin{center}} \def\ecc{\end{center}}     \def\npg{\newpage}
\def\beq{\begin{equation}} \def\eeq{\end{equation}} \def\hph{\hphantom}
\def\be{\begin{equation}} \def\ee{\end{equation}} \def\r#1{$^{[#1]}$}
\def\n{\noindent} \def\ni{\noindent} \def\pa{\parindent} 
\def\hs{\hskip} \def\vs{\vskip} \def\hf{\hfill} \def\ej{\vfill\eject} 
\def\cl{\centerline} \def\ob{\obeylines}  \def\ls{\leftskip}
\def\underbar#1{$\setbox0=\hbox{#1} \dp0=1.5pt \mathsurround=0pt
   \underline{\box0}$}   \def\ub{\underbar}    \def\ul{\underline} 
\def\f{\left} \def\g{\right} \def\e{{\rm e}} \def\o{\over} 
\def\vf{\varphi} \def\pl{\partial} \def\cov{{\rm cov}} \def\ch{{\rm ch}}
\def\la{\langle} \def\ra{\rangle} \def\EE{e$^+$e$^-$}
\def\bitz{\begin{itemize}} \def\eitz{\end{itemize}}
\def\btbl{\begin{tabular}} \def\etbl{\end{tabular}}
\def\btbb{\begin{tabbing}} \def\etbb{\end{tabbing}}
\def\beqar{\begin{eqnarray}} \def\eeqar{\end{eqnarray}}
\def\\{\hfill\break} \def\dit{\item{-}} \def\i{\item} 
\def\bbb{\begin{thebibliography}{9} \itemsep=-1mm}
\def\ebb{\end{thebibliography}} \def\bb{\bibitem}
\def\bpic{\begin{picture}(260,240)} \def\epic{\end{picture}}
\def\akgt{\noindent{Acknowledgements}}
\def\fgn{\noindent{\bf\large\bf Figure captions}}
\bd
\vskip-6.5cm
\hskip12cm{\large HZPP-9901}

\hskip12cm{\large Jan. 25, 1999}

\vskip2.5cm

\begin{center}
{\huge On the Intermittency and Chaos} 
\vskip0.5cm

{\huge in High Energy Collisions
\footnote{This work is supported in part by the National 
Natural Science Foundation of China.  (NSFC) under Grant No.19575021.}}
\vskip1.5cm

{Fu Jinghua \ \ \ \ \ \  Liu Lianshou\ \ \ \ \ \ \ Wu Yuanfang }
\vskip0.5cm

{\small Institute of Particle Physics, Huazhong Normal University, 
Wuhan 430079 China}
\vskip0.5cm

{\small Tel: 027 87673313 \qquad FAX: 027 87662646 
\qquad email: liuls@iopp.ccnu.edu.cn}
\date{ }
\begin{minipage}{125mm}
\vskip 2.5cm
\begin{center}{\Large Abstract}\end{center}
\ \ \ \   
It is shown that an event sample from the Monte Carlo simulation of a random
cascading $\alpha$ model with fixed dynamical fluctuation strength is 
intermittent but not chaotic,
while the variance of dynamical fluctuation strength in different events
will result in both the intermittency and the chaoticity behavior. 
This shows that fractality and chaoticity are two connected but different
features of non-linear dynamics in high energy collisions.
\end{minipage}
\end{center}
\newpage

\baselineskip 0.18in

For classical system, the description of non-linear behavior is
well established. It has been known by lattice calculation that the
classical non-Abelian gauge theory generally exhibits deterministic
chaos and that the Lyapunov exponent can be numerically
determined~\cite{classichaos}.  But for quantum system, because of the 
ambiguity associated with quantum chaos in the realm of quantization, 
nonconservation of the number of degrees of freedom and lack of 
a meaningful definition of a trajectory, there is no corresponding 
description existed. The study of non-linear behavior in high energy physics 
has, therefore, to be started phenomenologically. 

The first signal of such a behaviour came from 
the unexpectedly large local fluctuations in a single event of very high 
multiplicity recorded by the JACEE collaboration~\cite{JACEE}. Such large
fluctuations may not be simply due to statistical reason and was taken as
a signal of the existence of non-linear dynamical fluctuations.  It was soon 
realized that the idea can be applied to events of any multiplicity provided 
that a proper averaging of factorial moments is performed, as done in 
the pioneer work~\cite{BP} of Bia\l as and Peschanski. These authors have been 
able to show that, if the statistical fluctuations are of Bernouli (fixed 
multiplicity case) or Poisson (variable multiplicity case) type, the averaged 
factorial moments $F_q$ is equal to the averaged dynamical probability moments 
$C_q$. The anomalous scaling of the latter has taken the name of intermittency 
(or fractal). This led to extensive experimental studies~\cite{Kittel},
and the expected anomalous scaling has been observed successfully in the
experiments~\cite{NA2227}.

It should be realized, however, that the averaging procedure, apart from its
clear advantages, brings also a danger of losing some important information 
on the spatial patterns from event to event. In particular, some interesting 
effects, if present only in a part of events produced in high-energy collisions, 
may be missed. A possible example of this kind is the quark-gluon plasma which 
is expected to be characterized by specific intermittency exponents~\cite{QGP}. 
It seems therefore important and urgent~\cite{CaoHwa} to study the fluctuation 
of single-event moments $C_q^{(\rm e)}$ inside an event 
sample\footnote{It has been shown~\cite{BP} that the statistical fluctuations 
can be eliminated by using the factorial moments averaged over event sample. 
However, the extension of this method to single-event moments is highly 
non-trivial.  It is easy to show that the elimination of statistical 
fluctuations in single-event factorial moments $F_q^{(\rm e)}$ is incomplete. 
In order to avoid the complication caused by statistical 
fluctuations we will in this paper restrict ourself to the 
study of probability moments $C_q^{(\rm e)}$ directly.}. This fluctuation
is related to the chaotic behavior of the system~\cite{CaoHwa}. 
A quantity $\mu_q$ called entropy index can be introduced~\cite{CHPRD}
as an adequate parameter in measuring the chaotic behavior. The positivity of 
entropy index $\mu_q >0$ is proved to be a criterion for chaos~\cite{Sigma}.

Thus, two kind of non-linear phenomena --- fractality (intermittency) and 
chaoticity have been proposed in high energy collisions. In this short note 
we will study the relation beteen them using Monte Carlo simulation of a 
random cascading $\alpha$ model.  We will show that the anomalous scaling 
of the averaged probability moments (fractality or intermittency) and that 
of the event-space moments of single-event ones (chaoticity) are two connected 
but different features of non-linear dynamics. The system will exhibite both 
the fractal and the chaos behaviour only when the dynamical fluctuation 
strength is not fixed but is distributed over a certain range.

Let us first recall briefly the study of fractality (intermittency) in 
high energy collisions. This study is performed through the observation 
of anomalous scaling of averaged factorial moments $F_q$, which is
equal to the averaged probabilty moments $C_q$
\beq
F_q(M) = C_q(M) \equiv \frac{1}{M} \sum_{i=1}^M 
\frac{ \la p_i^q\ra} {\la p_i\ra^q} \propto M^{\phi_q},
\eeq 
where a phase space region $\Delta$ is divided into $M$ sub-cells, $p_i$
is the probability for a particle to fall in the $i$th sub-cell.

For a flat inclusive distribution the moment $C_q^{(\rm e)}$ for each event is 
defined as
\begin{equation}
C_q^{(\rm e)}= M^{q-1} \sum_{i=1}^M {\f(p_i^{(\rm e)}\g)}^q .
\end{equation}
We can now consider $C_q^{(\rm e)}$ not only through its average --- 
intended to get a better estimate of the hypothetical anomalous scaling of
single-bin moments, cf. eqn.(1) --- but also as a pattern-descriptor 
for particle fluctuations inside bins (just one among the many that could be 
devised). 

$C_q^{(\rm e)}$ may fluctuate greatly from event to event. In a sample
consisting of a large number $N$ of events, we get a distribution of 
$C_q^{(\rm e)}$, denoted by $P\f(C_q^{(\rm e)}\g)$, which is normalized to 
unity. The conventionally defined factorial moments, cf eqn.(1), give only 
an estimate of the mean of $P\f(C_q^{(\rm e)}\g)$. By taking the normalized 
moments of $P\f(C_q^{(\rm e)}\g)$ in event-space defined as
\begin{equation}
C_{p,q}=\la {C_q^{(e)}}^p\ra\f/\la C_q^{(e)}\ra^p\g. ,
\end{equation}
we have a quantification of the fluctuation of the spatial patterns, i.e. 
we can investigate the full shape of the distribution and, especially, the 
way it changes 
with the resolution $\delta=\Delta/M$. The value of $p$ can be any positive 
real number.  If $C_{p,q}(M)$ has a power law behaviour in $M$, i.e.
\begin{equation}
C_{p,q}(M) \propto M^{\psi_q(p)},
\end{equation}
then a new entropy index can be defined as,
\begin{equation}
\mu_q=\f.\frac{d}{dp}\psi_q(p)\g|_{p=1}.
\end{equation}

It is easy to see that finite, nonvanishing positive values of
$\mu_q$ corresponds to wide $P\f(C_q^{(\rm e)}\g)$, which in turn means 
unpredictable spatial pattern from event to event. By applying
the measure to known classical chaotic system, it has been 
shown~\cite{Sigma} that $\mu_q$ can be used as a measure of 
chaos in problems where only the spatial patterns can be 
observed and the positivity of $\mu_q$ is a criterion for chaos. 

An alternative way of calculating $\mu_q$~\cite{Sigma} is to express 
$C_{p,q}$ as
\begin{equation}
C_{p,q}=\la {\Phi_q^{(e)}}^p\ra,
\end{equation}
in which,
\begin{equation}
\Phi_q^{(e)}=C_q^{(e)}\f/\la C_q^{(e)}\ra\g.  .
\end{equation}
With the definition
\begin{eqnarray}
\label{e12}
\Sigma_q=\la \Phi_q^{(e)} \ln \Phi_q^{(e)} \ra,
\end{eqnarray}
we can obtain
\begin{equation}
\mu_q={\frac{\partial \Sigma_q}{\partial \ln M}}
\end{equation}
in the scaling region, i.e. where $\Sigma_q$ exhibits a linear 
dependence on $\ln M$. We will use this formula in calculating
the entropy indices $\mu_q$.

Let us turn now to the consideration of the relation between the fractality
and chaoticity in high energy collisions.  Since the random cascading 
$\alpha$-model~\cite{BP}\cite{KXTB} is often used to study the dynamical fluctuations 
in these collisions, we will use this simple model as a tool for our 
investigation. 

In the random cascading $\alpha$-model, the $M$ divisions of 
a phase space region $\Delta$ are made in steps. At the first step, 
it is divided into two equal parts; at the second step,  
each part in the first step is further divided into two equal parts, 
and so on. The steps are repeated until $M= {\Delta Y / \delta y}=2^{\nu}.$
How particles are distributed from step-to-step between the two
parts of a given phase space cell is defined by independent random 
variable $ \omega_{\nu j_{\nu}}$, where $j_{\nu}$ is the position
of the window ($1\le j_{\nu}\le 2^{\nu}$) and $\nu$ is the number of steps.
It is given by~\cite{KXTB}:
$$\omega_{\nu,2j-1}={1\over 2}(1+\alpha r) \ \ \ ; \ \ \ 
 \omega_{\nu,2j}={1\over 2}(1-\alpha r), \qquad j=1,\dots,2^{\nu-1}$$
where, $r$ is a random number distributed uniformly in the interval
$[-1,1]$. $\alpha$ is a positive number less than unity, which 
determines the region of the random variable $\omega$ and describes 
the strength of dynamical fluctuations in the model. After $\nu$ steps,
the probability in the $m$th window ($m=1,\dots,M$)
is $p_m=\omega_{1j_1}\omega_{2j_2}\dots 
\omega_{\nu j_{\nu}}$. Then according to eqn(1), probabilty moment 
$C_q^{(e)}$ in each event of different division steps are calculated,
and the moment $C_{p,q}$ and entropy index $\mu_q$ 
of the sample are obtained using eqn.(3) and eqn.(9).

Our research is done in the following two steps.

(A) Fix the model parameter to a definit value, say $\alpha=0.34$, the 
experimental results being around this value. The results of $\ln C_q$, 
$\ln C_{p,q}$ and $\Sigma_q$ vs. $\ln M$ from 6000 MC simulation events
are shown in Fig.1$(a),(b),(c)$ respectively. 

In Fig.1($a$) we see a straight line in bi-logarithm plot, which is an 
indication of intermittency (or fractality). However, the behavior of 
$\ln C_{p,q}$ vs. $\ln M$ in the model, cf. Fig.1($b$), is much different 
from the expected result for chaos~\cite{CaoHwa}. 
It does not show any scaling behavior or upward bending when $M$ goes 
larger as the chaotic behaviour requires~\cite{CaoHwa}. The first going 
up of $C_{p,q}$ is due to an intrinsic uncertainty of the intermittency
parameters~\cite{BiaZiaja}. The cascade responsible for intermittent behaviour 
has different realizations in different events, and the intermittency exponents 
determined from different realizations of the same random cascade 
are scattered around the average, i.e. the method has a finite 
resolution with respect to the parameters of the random cascade. The
$C_{p,q}$ saturates when $M$ goes large means that there isn't any essential
fluctuation of spatial pattern from event to event. Therefore,
this kind of $\alpha$ model cannot reflect the feature of chaoticity.
From the result showing in Fig.1($c$), using eqn.(9),
we can get for this case $\mu_q \sim 0$. If we donot consider
the finite resolution of model, there isn't any chaotic behavior.

In order to reproduce both intermittency (fractal) and chaos we take the second 
step.

(B) Instead of giving $\alpha$ a fixed value we let it be a
random variable having a Gaussian distribution. The mean value 
and variance of the Gaussian are both chosen as 0.22.
Calculating from 6306 events, the result of $\ln C_q$ vs. $\ln M$
are shown in Fig.2. It can be seen from the figure that there is very good 
power-law or scaling behavior, which means that though we have changed the 
method of setting model parameter, the anomalous scaling of the mean value of 
$C_q^{(e)}$ (intermittency phenomenon) survives. With the method developed in 
Ref.~\cite{strength} we can get the effective fluctuation strength in 
this case as $\alpha_{\rm eff}=0.337$. This value of $\alpha_{\rm eff}$ is 
within 
the limited range available in actual experiments. However, the behavior of 
$\ln C_{p,q}$ vs. $\ln M$ in the present case is much different from the case 
(A) and shows a typical behaviour of chaoticity, cf. Fig.3. 

From this result we see that a distribution of $\alpha$ will 
cause a distribution of single event probabilty moment $C_q^{(\rm e)}$,
i.e. for an event sample we have a wide $P(C_q^{(\rm e)})$. For increasing $M$,
along with the expected increase of the average, $P(C_q^{(\rm e)})$ will
show a rapid broadening, i.e. a more violent fluctuation of $C_q^{(\rm e)}(M)$
for different events, which will result in a even more unpredictable
spatial pattern from event to event. 

Using eqn.(8) and eqn.(9), we can 
calculate entropy indices for this case. The result of $\Sigma_q$ vs.
$\ln M$ are shown in Fig.4($a$). By performing a linear fit of $\Sigma_q$
vs. $\ln M$ in the range $M=8$ to $M=64$ (i.e. omitting the first three
points), $\mu_q$ is obtained and plotted in Fig.4($b$). 

By this two steps of MC simulation we can see that the procedure
of doing simulation with random cascading $\alpha$ model of fixed strength
parameter $\alpha$, as has been widely used before, captured only 
one aspect of the non-linear property (intermittency) but cannot reproduce 
the fluctuation of spatial patterns from event to event.  It will cause 
the losing of information on the spatial patterns in different events 
and some intersting effects if they are present only in a part of events,
may be lost. If we want to give a more complete description of the non-linear 
properties using $\alpha$ model, the model parameter cannot be fixed. 

In conclusion, the nonvanishing positive values of $\mu_q$, which is an 
indication of chaos, correspond to wide $P(C_q^{(\rm e)})$, which in turn means 
unpredictable spatial pattern from event to event. Such an unpredictability 
of wide $P(C_q^{(\rm e)})$ is caused by different dynamical fluctuation strength 
in different events, i.e. by a distribution of dynamical fluctuation strength 
in an event sample. Events in one sample are all beginning with
similar initial condition. During the collision process each event
will evolve with a different strength of dynamical fluctuation
and result in a fluctuation of spatial pattern in final state event
space. 

Dynamical fluctuation strength is directly related to the dynamical
mechanism in a particular collision. We take the distribution of 
dynamical fluctuation strength, i.e. the distribution of model parameter
$\alpha$, to be a Gaussian only because it is the most common distribution 
of random variables in nature. We have also tried a uniform distribution
of $\alpha$ and non-zero positive entropy indices $\mu_q$ can be obtained too. 
(The result is not shown here). Revealing the distribution
of dynamical fluctuation strength in different collisions will be
a very constructive work and it will certainly help us a lot in studying
the mechanism of strong interactions. How the different distributions 
of dynamical fluctuation strength together with it's mean value and width
will influence the entropy index of an event sample is also a problem 
worthwhile further investigation.

As has been stressed in the introduction, this study is restricted to the
probability moments and the problem of eliminating statistical 
fluctuations in experimental data analysis has been postponed. To develope
an effective method for eliminating the statistical fluctuations for
single-event moments is a challenge for future investigation.

\newpage
\def\Journal#1#2#3#4{{#1} {\bf #2} (#3) #4}
\def\NCA{\em Nuovo Cimento} \def\NIM{\em Nucl. Instrum. Methods}
\def\NIMA{{\em Nucl. Instrum. Methods} A} \def\NPB{{\em Nucl. Phys.} B}
\def\PLB{{\em Phys. Lett.}  B} \def\PRL{\em Phys. Rev. Lett.}
\def\PRD{{\em Phys. Rev.} D} \def\ZPC{{\em Z. Phys.} C}
\def\PRE{{\em Phys. Rev.} E} \def\PRC{{\em Phys. Rev.} C} 

\vskip1cm
\ni
{\Large\bf Figure Captions}
\vskip0.8cm

\ni
{\bf Fig.1} \ ($a$) Averaged $C_2$, ($b$) $C_{p.2}$, ($c$) $\Sigma_2$ 
for fixed $\alpha$.  Full lines are for guiding the eye.
\vskip0.5cm

\ni
{\bf Fig.2} \ Averaged $C_2$ for Gaussian-distributed $\alpha$.
\vskip0.5cm

\ni
{\bf Fig.3} \ \ The ln$C_{p,q}$ vs. ln$M$ for 
the random cascading model with 

\qquad Gaussian-distributed $\alpha$. Full lines are for guiding the eye.

\vskip0.5cm
\ni
{\bf Fig.4} \ $\Sigma_q$ and $\mu_q$ for Gaussian-distributed $\alpha$.  
Full lines are for guiding the eye.

\newpage

\begin{picture} (260,240) 
\put(-120,-400)   
{\epsfig{file=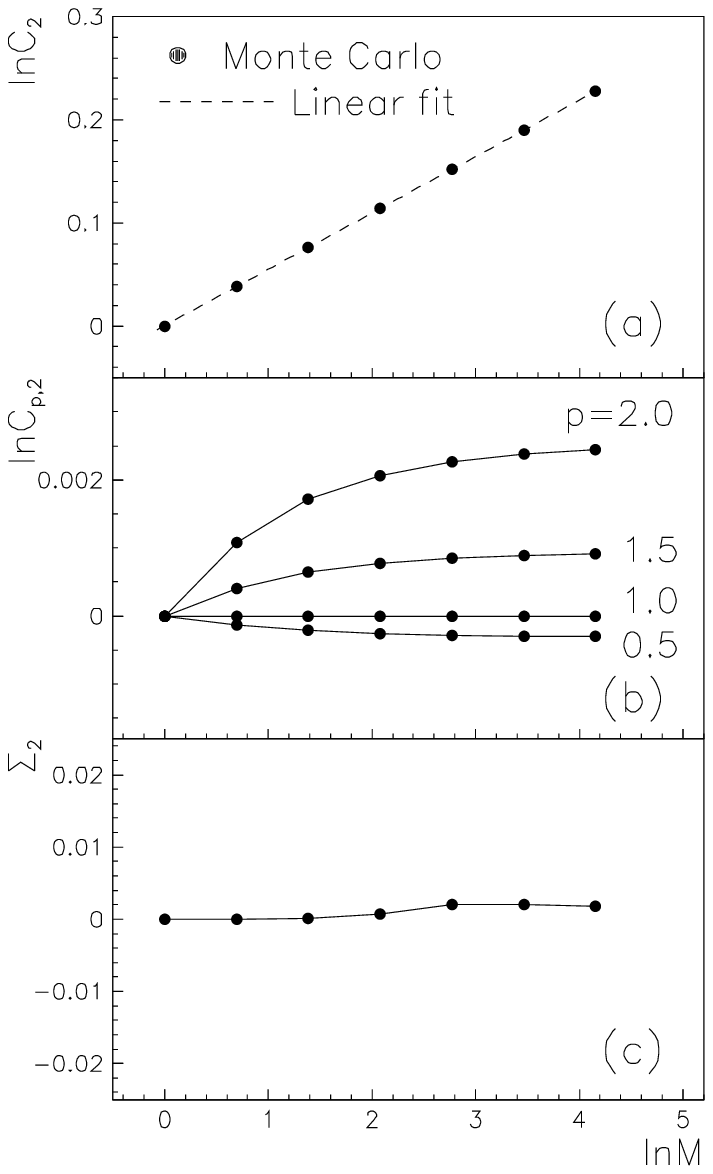,bbllx=0cm,bblly=0cm,
           bburx=8cm,bbury=6cm}}  
\end{picture}
 \vs7.5cm

\cl{{\bf Fig.1} \ ($a$) Averaged $C_2$, ($b$) $C_{p.2}$, ($c$) $\Sigma_2$ 
for fixed $\alpha$.}
\cl{ Full lines are for guiding the eye.}

\newpage

\begin{picture} (260,240) 
\put(-120,-400)   
{\epsfig{file=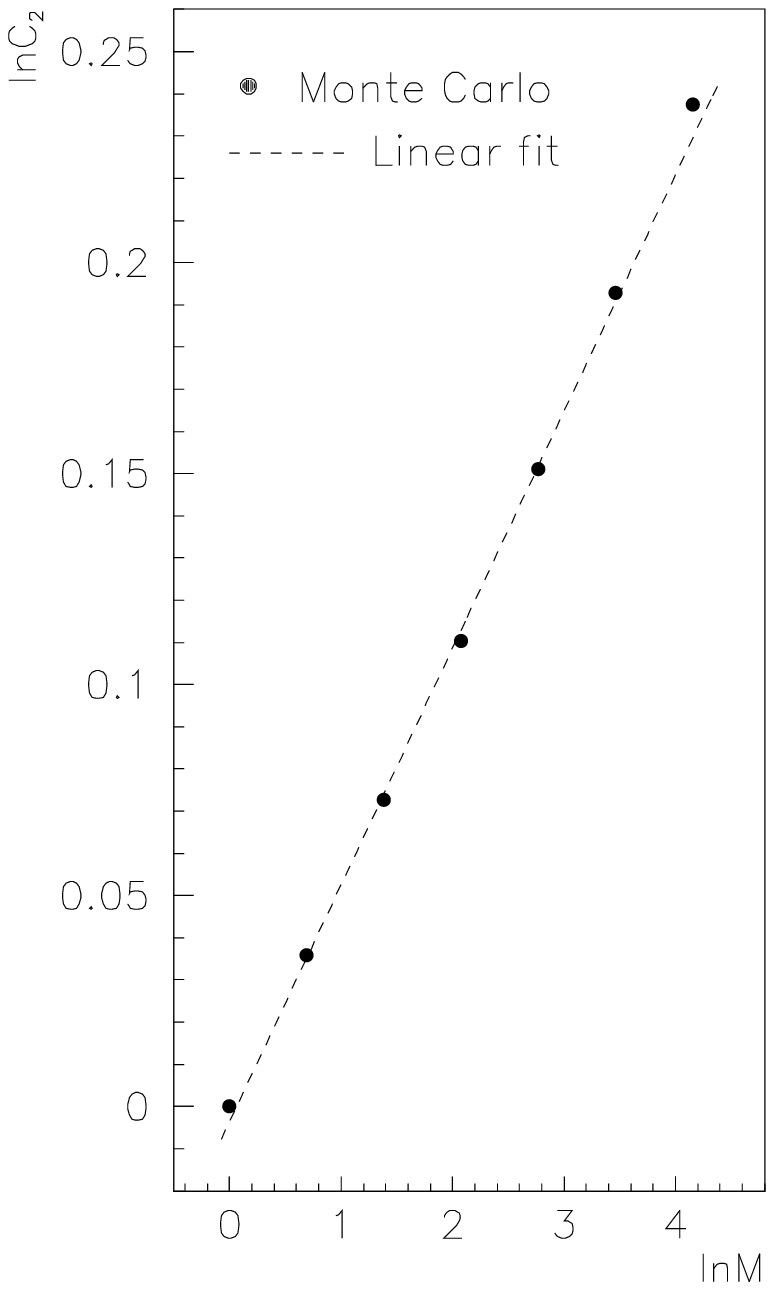,bbllx=0cm,bblly=0cm,
           bburx=8cm,bbury=6cm}}  
\end{picture}
\vskip8.0cm

\cl{{\bf Fig.2} \ Averaged $C_2$ for Gaussian-distributed $\alpha$.}

\newpage

\begin{picture}(260,240)
\put(-108,-250)        
{\epsfig{file=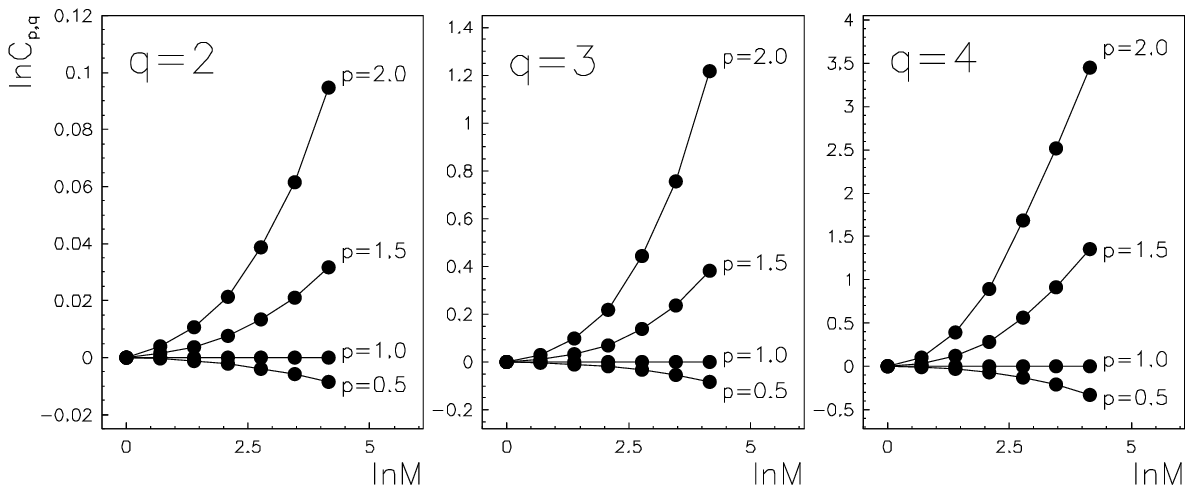,bbllx=0cm,bblly=0cm,
	   bburx=8cm,bbury=6cm}}
\end{picture}

\vs-2.0cm
\cl{{\bf Fig.3} \ \ The ln$C_{p,q}$ vs. ln$M$  for
the random cascading model with}
\cl{Gaussian-distributed $\alpha$. Full lines are for guiding the eye.} 

\begin{picture} (260,240) 
\put(-108,-350)        
{\epsfig{file=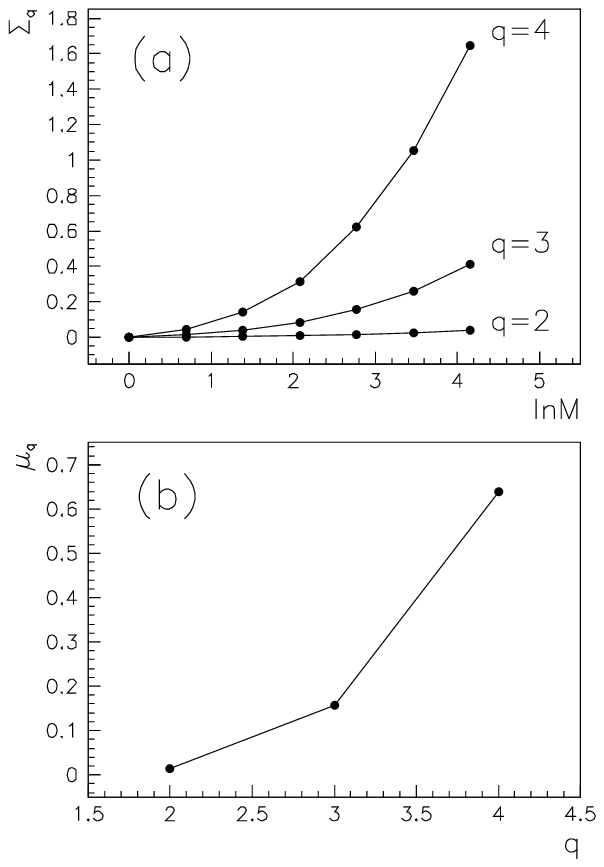,bbllx=0cm,bblly=0cm,
	   bburx=8cm,bbury=6cm}}
\end{picture}
\vs3.0cm
\cl{{\bf Fig.4} \ $\Sigma_q$ and $\mu_q$ for Gaussian-distributed $\alpha$.}
\cl{ Full lines are for guiding the eye.}

\ed